\documentclass[prl,aps,floatfix,twocolumn]{revtex4}
\usepackage{latexsym,amssymb}
\usepackage{graphicx}% Include figure files
\usepackage{subfigure}
\usepackage{amsmath, amsthm, amssymb}

\begin{document}

\preprint{}

\title{Shapes of Semiflexible Polymers in Confined Spaces}

\author{Ya Liu, Bulbul Chakraborty}

\affiliation{Martin Fisher School of Physics, Brandeis University, Mailstop 057, Waltham, Massachusetts 02454-9110, USA}
\date{\today}

\begin{abstract}

We investigate the conformations  of a semiflexible polymer confined to a square box. Results of Monte Carlo simulations
show the existence of a shape transition when the persistence length of the polymer becomes comparable to the dimensions
of box. An order parameter is introduced to quantify this  behavior. A simple mean-field model is constructed to study
the effect of the shape transition on the effective persistence length of the polymer.

\end{abstract}

%\pacs{PACS numbers: 61.20.Lc, 64.70.Pf, 64.60.Cn}

\maketitle

\section{Introduction}
%%BC
Biological macromolecules live in the crowded, confined environment of a cell.   They are often packed into spaces that
are much smaller than their natural size and adopt conformations which are unlikely to occur in free space.  For
example, viral DNA is packaged in a capsid whose dimensions are comparable to the persistence length of DNA
\cite{Gelbart,Odijk01,Yeomans01,Jiang,Katzav}, which obliges the DNA to adopt a tightly bent shape. Other examples of
strong confinement include actin filaments in eukaryotic cell \cite{Sarah}, and protein encapsulated in E.Coli
\cite{Marrison}.
%Closely packing polymer in certain geometry is a crucial phenomenon%
%in biology and mechanics. Examples include the viral
%DNA packaging in the capsids\cite{Yeomans01,Jiang,Katzav},actin filaments confined in eukaryotic cell \cite{Sarah}, and
%the protein encapsulated in the E.Coli\cite{Marrison}. 
In such environments, the macromolecule is forced to optimize the competing demands of configurational entropy, excluded
volume and bending energy.  The packaging of DNA in viral capsids has been investigated extensively and illustrates the
effects of these competing interactions and the additional electrostatic forces arising from its charged nature
\cite{Gelbart,Odijk01}. Experimental and theoretical studies \cite{Yeomans01,Muthukumar01,Frenkel} show that the static,
and dynamic properties such as threshold force for packaging and packaging time,  change drastically with changes in
confining geometry.  Another example is provided by driving DNA through a nanochannel by electrical force
\cite{Clementi,Kolomeisky}, where the dimensions of the channel determine the translocation time and threshold voltage. 
 Understanding the statistics of polymer in confinement is therefore essential for exploring  both the static 
and dynamic properties of DNA \cite{Sakaue}.

In this paper, we investigate the shape and rigidity of a single, semiflexible polymer under confinement. Using the
Bond-Fluctuation-Model (BFM) \cite{Kremer01}, we simulate a single, semiflexible  polymer chain confined to a box in two
dimensions. 
One advantage of  the two-dimensional (2D) simulations is that they can be compared directly to experimental
measurements, 
since conformations of DNA on a substrate can be visualized through atomic force microscopy (AFM) which characterize the
statistics of fluctuations of DNA\cite{Arneodo}. Our simulations show that there is a shape transition as a function of
increasing rigidity.  We analyze the nature of this transition by defining an order parameter associated with the
transition.  As expected, the flexibility of the chain, measured by the tangent-tangent correlation function, evolves
with the shape change.  We construct a theory for this correlation function by considering Gaussian fluctuations around
the ``average shape''.

\section{Simulations} 
\begin {figure}
 \centering
 \includegraphics[width=8cm]{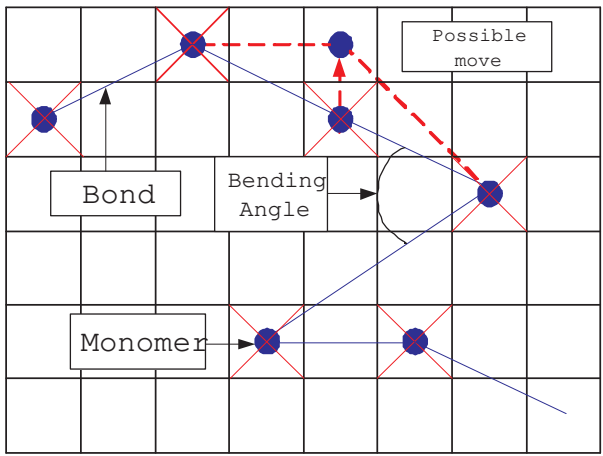}
\caption{One configuration of polymer in BFM. The bond vectors are restricted to the set B (Eq. \ref{bondset}) and one
possible move is denoted by the dashed line.}
 \label{BFM}
 \end {figure}

The simulations are performed using the bond fluctuation algorithm, introduced by Kremer and
Carmersin \cite{Kremer01}. The bond fluctuation model (BFM) is a coarse-grained model of a polymer in which the chain
lives on a hypercubic lattice and fluctuations on scales smaller than the lattice constant are suppressed. 
The polymer is represented by a chain of effective monomers connected by effective bonds.  The effective bonds and
effective monomers are constructed to account for excluded volume effects. In 2D, each monomer occupies an elementary
square and forbids other monomers to occupy its nearest and next-nearest neighbors \cite{Binder-review}.  
%find the reference
In order to avoid crossing of bonds, the effective bonds are restricted to the set:

\begin{align}
 B = P(2,0)\cup P(2,1)\cup P(2,2)\cup P(3,0) \nonumber\\
     \cup P(3,1) \cup P(3,2)
\label{bondset}
\end{align} in units of lattice spacing, and $P(m,n)$ stands for the sign combinations $\pm m, \pm n$ and all the
permutations. With this restriction, the average bond length (Kuhn length b) is 2.8. Fig. \ref{BFM} illustrates the BFM
representation of a polymer and shows one allowed move that obeys the self-avoidance condition.

The only non-bonded interaction included in this representation is excluded volume.  To represent a semiflexible
polymer, such as DNA in this model, the energy cost of bending has to be incorporated. The mechanical properties of DNA
are well described by the worm-like chain model (WLC) \cite{Yamakawa}, in which DNA is characterized by its total
contour length $L$ and persistence length $l_{p}$.  
In the continuum limit, the bending energy $H_{b}$ in D-dimensional space can be expressed as:
\begin{eqnarray}
 H_{b} =\frac{\kappa}{2} \int_{0}^{L} \bigl[\frac{d\bold{u}(s)}{ds}\bigr]^2\,ds
\end{eqnarray}
where the stiffness, $\kappa$, is related to $l_{p}$ by $\frac{\kappa}{k_{B}T}=\frac{D-1}{2}l_{p}$, and 
$l_{p}$ characterizes the decay length of the correlation between the tangent vectors \cite{Yamakawa}
\begin{eqnarray}
\langle\bold{u}(s)\cdot\bold{u}(0)\rangle\propto e^{-\frac{s}{l_{p}}} ~.
\end{eqnarray}
Here  $\bold{u}(s)=
\partial\bold{R}(s)/
\partial{s}$ 
is the tangent vector at arclength s, and $\bold{R}(s)$ is the position vector. The effect of confinement on the form of
this correlation function is one of the questions we explore in this paper. In the following, all energies will be
measured in units of $k_{B}T$.

In a lattice representation such as the BFM,  the bending energy $H_{b}$ associated with a  configuration is expressed
as \cite{Erwin}: 
$$
H_{b}=
\frac{l_{p}}{2b}\sum_{i=1}^{N-1}\bold{u}_{i}\cdot\bold{u}_{i+1} $$
where $N$ is the total number of monomers in the chain. 
The modified BFM model now includes excluded volume interactions and bending rigidity. Monte Carlo calculations are
carried out using the Metropolis Algorithm with  $H_{b}$  as the energy function since excluded volume effects are
incorporated in the allowed configurations of the BFM.

\begin {figure}
\begin{center}
\subfigure[] {\label{setupa}
 \includegraphics[width=7.5cm,height=7.5cm]{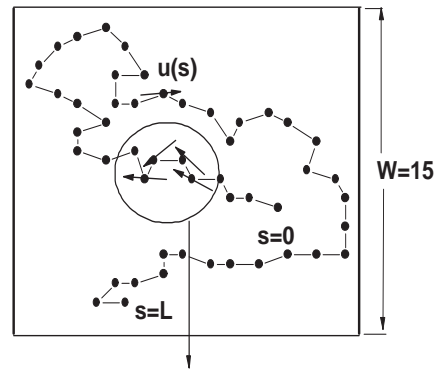} }

\subfigure[] { \label{setupb}
  \includegraphics[width=6.5cm,height=5cm]{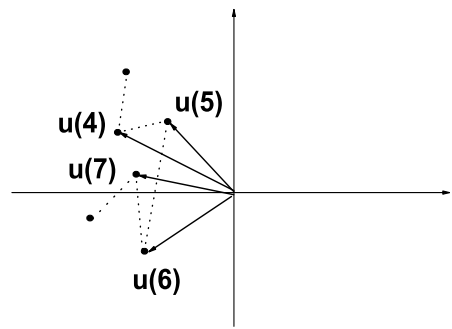} }
 \caption{$(a)$ One initial configuration of our simulation. The contour length of the chain is 60, box size is 15.
$\bold{u}(s)$ is illustrated by arrows. $(b)$ Mapping part of the chain circled in $(a)$ onto $\bold{u}(s)$ space.}
 \label{Polymer in box}
\end{center}
 \end {figure}

Fig. \ref{Polymer in box} illustrates the simulation setup for a semiflexible chain in a square box with linear
dimension $W$, in which  all lengths are measured in units of the Kuhn length. The radius of gyration $R_{g}$ of the
unconfined semiflexible chain is proportional to $L^{\nu}l_{p}^{1-\nu}$ \cite{Ya}, where $\nu$ is the Flory exponent
($\nu=\frac{3}{4}$ for 2D, $\nu\approx 0.588$ for 3D \cite{Doi}) . We set up conditions such that $R_{g}$ is larger than
the box size. For a chain of contour length $L=60$, $W$ has to be less than $21.5$, for this condition to be met.  Most
of our results were obtained for $L=60$ and $W=15$, however, limited sets of data were also collected for $L=80$, $W=15$
and $L=40$, $W=10$ to investigate the effects of $L$ and $W$.  In addition, we compare the results of a circular box to
a square box. Each Monte Carlo trajectory is allowed to equilibrate for $10^9$ Monte Carlo steps which correspond to
many ($\sim100$) Rouse relaxation times \cite{Doi}. The range of the persistence length $l_{p}$ is varied from 0,
corresponding to a flexible chain, up to $W$, which is the rod-like limit. For each $l_{p}$, more than 20 independent
runs were performed to get adequate statistics. 

\section{Shapes in confinement} 
For a semiflexible chain, the mechanical properties are well described by a self-avoiding walk \cite{Valle}, and the
only effect of the semiflexibility is to change the Kuhn length \cite{Ya}.   In confinement, especially if the confining
dimensions are smaller than or comparable to the persistence length,  semiflexibility plays a much more important role
since the bending energy is in intense competition with the entropy.  As a result, the conformations adopt an average
shape which is influenced by the confining geometry \cite{Yeomans01,Katzav,Chen,Yuan}.   

\begin{figure}
 \centering
\subfigure[] {
\label{lp1a}
\includegraphics[width=7cm,height=5.5cm]{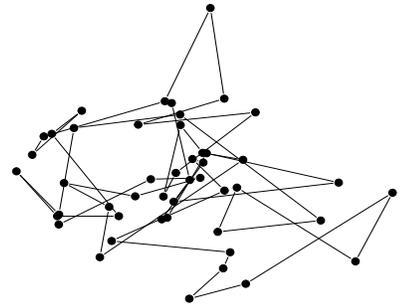} }
\subfigure[] {
\label{lp5b}
\includegraphics[width=7cm,height=5.5cm]{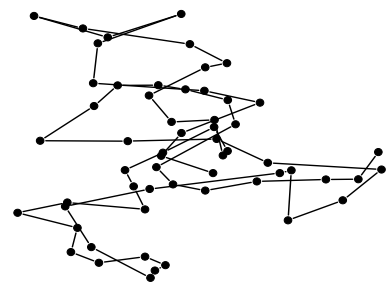} }
\caption{$(a)$ Plot of $\langle\bold{u}(s)\rangle$ using the mapping illustrated in Fig. \ref{setupb} for a polymer of
$l_{p}=1$. $(b)$ $l_{p}=5$.}
 \label{lpsmall}
\end{figure}

We visualize and classify a chain's configurations using the 
average tangent vector, $\langle\bold{u}(s)\rangle$, along the chain. The chain configuration is mapped onto the tangent
vector space by calculating the tangent vector at each monomer, and then connecting the end points of these vectors
according to their chain sequence, as shown in Fig. \ref{setupb}. The average $\langle\bold{u}(s)\rangle$ can then be
calculated by averaging over Monte Carlo trajectories. We choose $\langle\bold{u}(s)\rangle$ instead of position vector
to characterize the shape because the former is independent of the starting point of the chain.  We average over $10^5$
statistically independent configurations generated by a Monte Carlo trajectory. Fig. \ref{lpsmall} shows
$\langle\bold{u}(s)\rangle$ for $l_{p}=1$ and $l_{p}=5$. These two shapes look very similar: tangent vectors are
arranged randomly so that the connecting bonds cross each other frequently.  This is not surprising since for $l_{p} <<
W,L$, the conformations are close to that of a self-avoiding walk and there is no ``order'' in the tangent vector space.

\begin{figure}
 \centering
\subfigure[] {
\label{lp15}
\includegraphics[width=7cm,height=5.5cm]{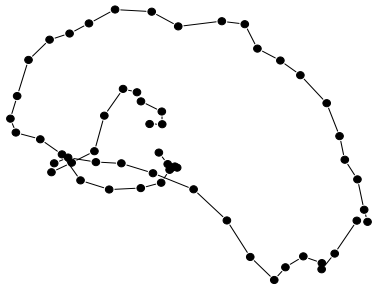}
 }
\subfigure[] {
\label{lp25}
\includegraphics[width=7cm,height=6.5cm]{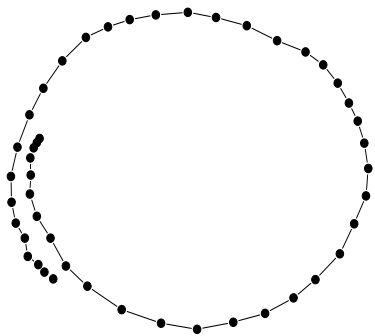} }
\caption{$(a)$ Same as Fig. \ref{lpsmall} but for $l_{p}=15$. $(b)$ $l_{p}=25$. }
 \label{lplarge}
\end{figure} 
The situation is dramatically different for $l_{p} \gtrsim W$ as illustrated in Fig. \ref{lplarge}. For
$5<l_{p}<15$, there are vestiges of order in $\bold{u}(s)$ space but for $l_{p} \geq 15$, the configurations demonstrate
clear signatures of ordering as measured by $\langle\bold{u}(s)\rangle$ (Fig. \ref{lplarge}). At the onset of ordering
around $l_{p} \simeq W$, there are extended correlations between the tangent
vectors but the shape is  more like an ellipse than a circle.  We believe that the reason behind this is the
square shape of the box.  In the box, the diagonal direction offers maximal space, therefore when the bending energy
first starts seriously competing with the chain entropy, an anisotropic shape can better optimize the entropy than an
isotropic one.  
To check the validity of this idea, we analyzed the shape of a chain confined in a circle with similar parameters:
$L=60$, $l_{p}=15$ and a diameter of 15. In this case, $\langle\bold{u}(s)\rangle$ was found to be isotropic.   For
$l_{p}$ much larger than the box size there is a clear spiral shape to the chain (Fig. \ref{lp25}).   %%BC put different
%labels in the figures so that you can refer to these and not the Fig numbers explicitly.

\begin{figure}
 \centering
\includegraphics[width=7cm,height=5.5cm]{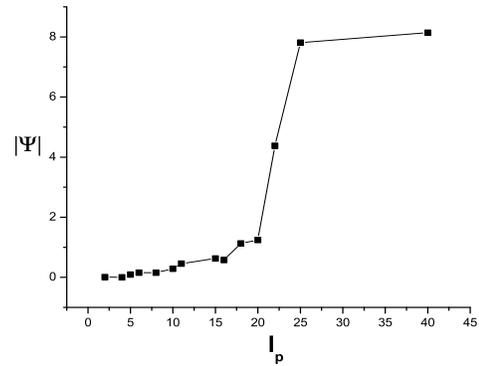}
\caption{Absolute value of the order parameter, $\vert\Psi\vert$, as a function of $l_{p}$.}
\label{orderparam}
\end{figure}

In order to better quantify the order in the shapes and to explore the nature of the transition from a state with no
visible ordering to the clear spiral shapes for $l_{p} \ge 15$, we define an order parameter, $\Psi$ associated with the
tangent vector:
\begin{equation}
\label{orderparameter}
 \Psi=\langle\int_{0}^{L}\bold{u}(s)\times\frac{
\partial\bold{u}(s)}{
\partial s}\,ds\rangle
\end{equation}
Since the chains live in 2D,  $\Psi$ can be considered as a scalar order parameter.   From geometry \cite{Novikov}, the
direction of $\bold{u}(s)\times\frac{
\partial\bold{u}(s)}{
\partial s}$ is along the binormal vector and its  magnitude is the curvature. Therefore $\Psi$ is zero for a shape with
no preferred bending direction such as an undulating line but is nonzero for a shape such as a spiral. Measuring
$\vert\Psi\vert$ as a function of $l_{p}$ for a given $L$ and $W$, we find a sharp change at $l_{p}\sim 15$ which is
close to the box size W, as shown in Fig. \ref{orderparam}.

To analyze the nature of this transition, we measured the distribution of the order parameter at different points along
the transition.%%BC   there is a  l_{b} and l_{p} and l_{{e}} .  Make sure these are being used correctly
\begin{figure}
 \centering
\includegraphics[width=7cm,height=5.5cm]{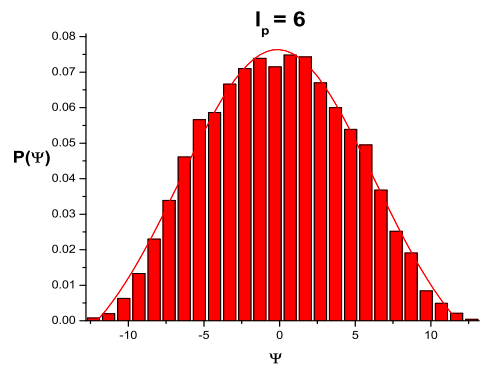}
 \caption{Distribution function $P(\Psi)$ for $l_{p}=5$. The red line denotes a Gaussian fit. The mean is at $\Psi=0$
and the configurations are disordered (Fig. \ref{lpsmall}).}
\label{Distributionlp5}
\end{figure}

\begin{figure}
 \centering
\includegraphics[width=7cm,height=5.5cm]{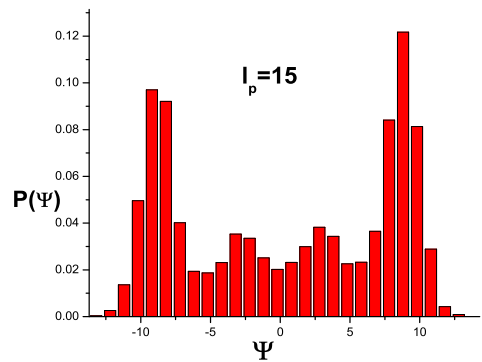}

 \caption{Distribution function $P(\Psi)$ for $l_{p}=15$. There are 4 peaks and a valley at $\Psi=0$, and this is no
longer described by a Gaussian.}
 \label{Distributionlp15}
\end{figure}

\begin{figure}
 \centering
\includegraphics[width=7cm,height=5.5cm]{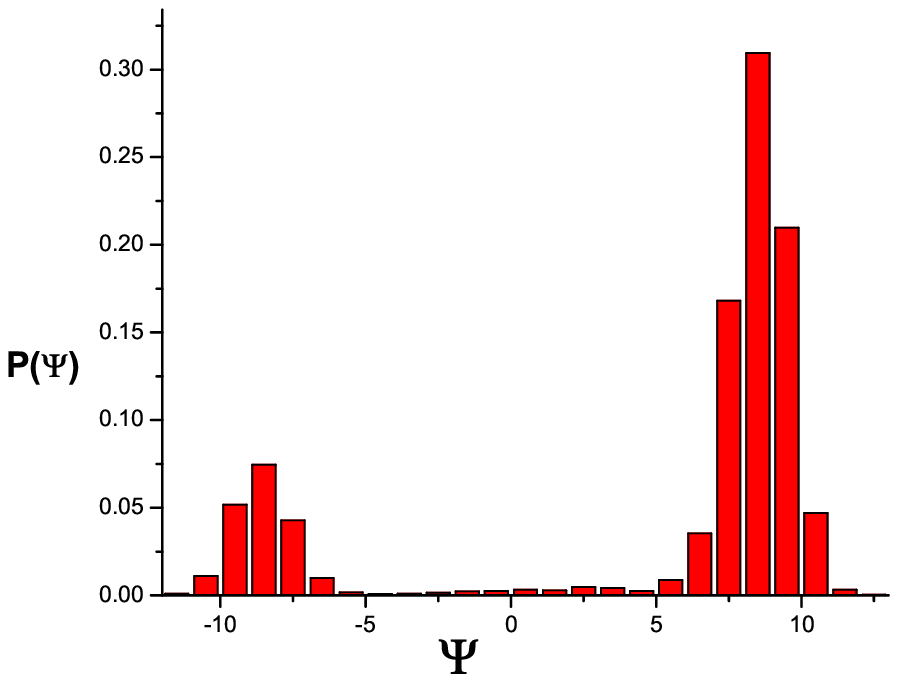}

 \caption{Distribution function $P(\Psi)$ for $l_{p}=25$.}
 \label{Distributionlp25}
\end{figure}

At $l_{p}=5$,  the distribution is close to a Gaussian, as seen in Fig. \ref{Distributionlp5}.  
With increasing values of $l_{p}$ the distribution becomes significantly non-Gaussian (Fig. \ref{Distributionlp15}) and
ultimately evolves to a bimodal distribution resembling two Gaussians(Fig. \ref{Distributionlp25}). The distributions
are reminiscent of a first order transition, except for the fact that for $l_{p} \simeq W$, the distributions clearly
show four peaks as evident in Fig. \ref{Distributionlp15}.

\begin{figure}
 \centering
\includegraphics[width=7cm,height=5.5cm]{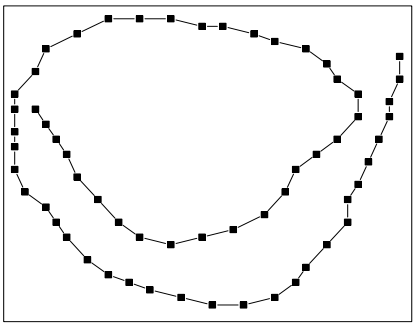}

 \caption{A configuration in the square box corresponding to the outer peak in Fig. \ref{Distributionlp15}.}
 \label{Outerpeak}
\end{figure}

\begin{figure}
\centering
\subfigure[] {
\label{innerpeakconfig}
\includegraphics[width=7cm,height=5.5cm]{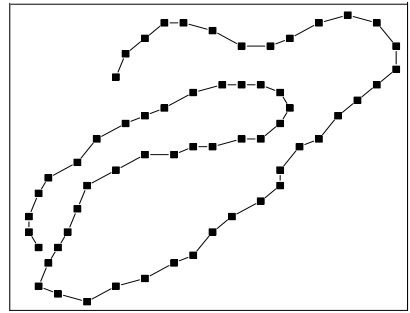} }
\subfigure[] {
\label{circleinnerpeak}
\includegraphics[width=6.5cm,height=5.5cm]{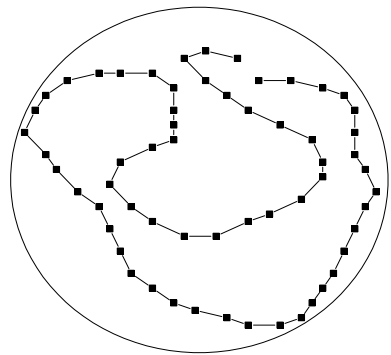} }
 \caption{$(a)$ A configuration in a square box corresponding to the inner peak in Fig.\ref{Distributionlp15}. $(b)$
Same as $(a)$ but in a circular box.}
\label{Peakconfig}
\end{figure}

Typical configurations corresponding to the outer peaks are similar to the plot in Fig. \ref{Outerpeak};  spiral in a
box.  A typical configuration for an inner peak  is illustrated in Fig. \ref{innerpeakconfig} and shows an elliptical
character. In a circular box, the inner peaks are still evident but they do not have the elliptical character (Fig.
\ref{circleinnerpeak}).   These peaks are, therefore, more robust than the elliptical shape.  Since the only other
length scale in the problem, besides, $l_{p}$ and $W$  is the contour length or, equivalently, $R_{g}$, we have future
plans of exploring the distributions as a function of $R_{g}$  in order to better understand this intermediate state.

Interactions of a semiflexible polymer with hard (or semisoft) boundaries are known to introduce orientational
correlations \cite{Katzav,Sakaue1,Yuan,Chen}. One example is the transition to liquid crystalline phases of polymers
confined between two hard walls \cite{Yuan} when the persistence length is comparable to the dimensions of the
confinement. The shape changes that we are observing in our study have a similar physical origin.

\section{Tangent-tangent correlations and effective persistent length in confinement} 
\begin {figure}
 \centering
 \includegraphics[width=7cm,height=5.5cm]{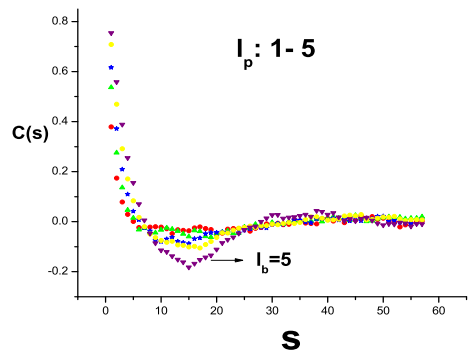}

 \caption{$C(s)$ (Eq.$\ref{C(s)}$) for $l_{p}\leq 5$. When $l_{p}< 5$, $C(s)$ decays exponentially and then fluctuates
around 0; a behavior that is similar to an unconfined chain. When $l_{p} = 5$, $C(s)$ begins to show the oscillatory
structure. The meansured persistence length $l_{e} \sim l_{p}$.}
 \label{correlation5}
 \end {figure}

\begin {figure}
 \centering
 \includegraphics[width=7cm,height=5.5cm]{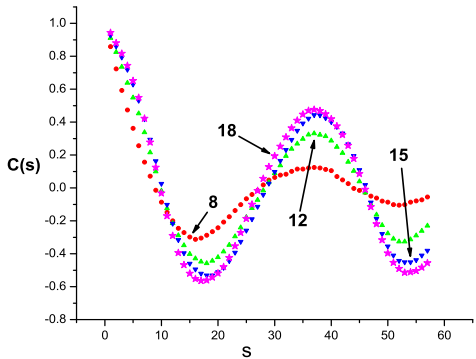}

 \caption{C(s) (Eq.$\ref{C(s)}$) for $8\leq l_{p}\leq 18$. Graphs are labelled by the value of$l_{p}$. In contrast to
the behavior observed at small values of $l_{p}$ (\ref{correlation5}), there is a distinct periodic behavior. }
 \label{correlation18}
 \end {figure}

>From the discussion above, it is
clear that the shape, characterized by the average tangent vector, depends on the relation between  $l_p$ and $W$. The
tangent-tangent correlation function, $C(s,s^{\prime})$, defined as 

\begin{eqnarray}
\label{C(s)}
C(s,s^{\prime})\equiv\langle(\bold{u}(s)-\langle\bold{u}(s)\rangle)\cdot(\bold{u}(s^{\prime})-\langle\bold{u}(s^{%
\prime})\rangle)\rangle
\end{eqnarray}
is expected to depend on the shape of the polymer. Such changes have been observed in experiments. For example, in actin
filaments trapped in narrow channels, $C(s,s^{\prime})$ exhibits a oscillatory behavior, and an effective persistence
length, deduced from the correlation function,  shows changes from the bare persistence length \cite{Sarah,Odijk}.
Because of translational invariance and the indistinguishability of two ends of a polymer chain, $C(s,s^{\prime})$ is
only a function of $\vert s-s^{\prime} \vert$, and in the simulations, we measure 
$C(s)\equiv C(s,0)$. Figs. \ref{correlation5}, and \ref{correlation18} show $C(s)$ measured at $l_{p}\leq 5$ and $8\leq
l_{p}\leq 18$, respectively. 

For $l_{p}< 5$, $C(s)$ exhibits exponential decay with a length scale which is indistinguishable from the bare
persistence length, $l_{p}$. 
In this regime, where $l_{p}$ is much smaller than $W$, the free energy of the polymer is dominated by the entropy , and
the fluctuations resemble those of an unconfined semiflexible polymer.

\begin {figure}
 \centering
 \includegraphics[width=8cm]{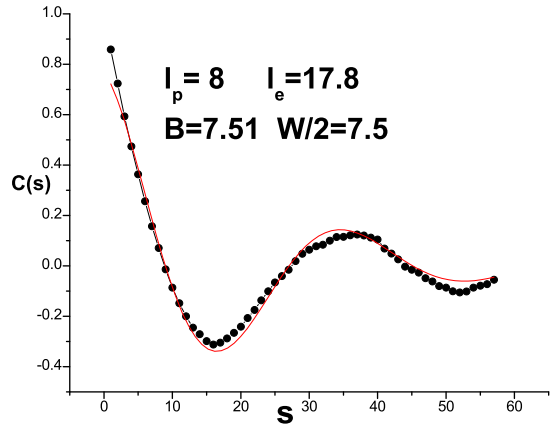}

 \caption{Fit of C(s) by the function $Ae^{-\frac{s}{l_{e}}}\cos\frac{s}{B}$ for $l_{p}= 8$. Box size $W=15$, and $B
\simeq\frac{W}{2}$.}
 \label{Fitting8}
 \end {figure}

\begin {figure}
 \centering
 \includegraphics[width=7cm,height=5.5cm]{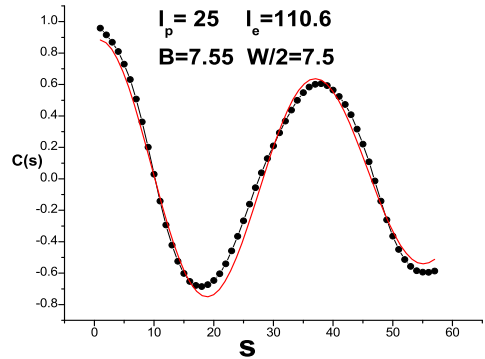}

 \caption{Fit of C(s) by the function $Ae^{-\frac{s}{l_{e}}}\cos\frac{s}{B}$ for $l_{p}= 25$.$l_{e}$ changes and $B
\simeq\frac{W}{2}$.}
 \label{Fitting25}
 \end {figure}

\begin {figure}
 \centering
 \includegraphics[width=7cm,height=5.5cm]{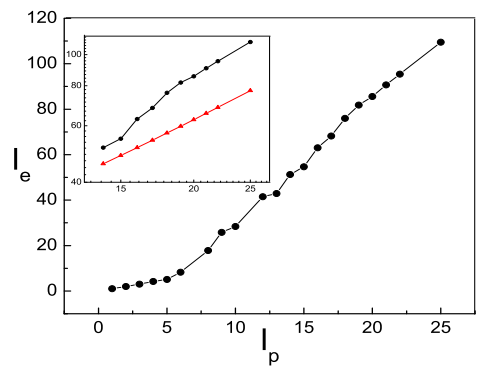}

 \caption{Plot of the effective persistence length $l_e$ versus the bare persistence length $l_p$ for $W = 15$, $L =
60$. In the inset, the black dots represent $l_e$ vs. $l_p$ for $l_p \geq 13$ and the red triangles represent the plot
of the function (\ref{leff}).}
 \label{le-lp}
 \end {figure}

The situation changes for $l_{p}$ larger than or comparable to $W$. As seen from Fig. \ref{correlation18}, $C(s)$
exhibits an oscillatory character.  
A negative value of $C(s)$ indicates a reflection in orientation of the tangent vector.  The oscillations, therefore, 
reflect the constraints imposed on $C(s)$ by the geometry of the confinement and we expect the periodicity to be related
to the size of the box, $W$. We therefore represent $C(s)$ as $Ae^{-\frac{s}{l_{e}}}\cos\frac{s}{B}$, with A,B and
$l_{e}$ as fitting parameters. From Fig. \ref{Fitting8} and \ref{Fitting25}, we see that this form captures the
properties of $C(s)$.  In the regime $l_{p}\geq 8$,  the parameter $B$ is not sensitive to $l_{p}$ and approaches the
value 7.5, which is equal to $W/2$. 
We can extract  the effective persistence length, $l_{e}$,  from the fits. The variation of this length, which
characterizes the rigidity of the polymer under 
confinement is shown in Fig. \ref{le-lp}. It is clear that $l_{e}$ begins to deviate from $l_{p}$ for $l_{p}\gtrsim 8$.
This is the regime in which the distribution of the order parameter $\Psi$ starts exhibiting multiple peaks but the
average order parameter is still close to zero. 

\section{Gaussian theory of  $C(s)$} 
\begin {figure}
 \centering
 \includegraphics[width=7.5cm,height=7cm]{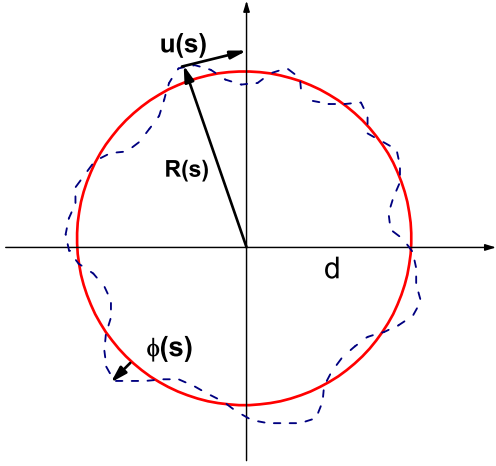}

 \caption{Mean field model for calculating tangent-tangent correlation function.}
 \label{Meanfieldmodel}
 \end {figure}

To understand the origin of the changes in $C(s)$ with the evolving shape of the polymer, we construct a Gaussian model
of the fluctuations of the tangent vector. In this model (Fig. \ref{Meanfieldmodel}), the polymer 
fluctuates around a circle with radius $d$, and the magnitude of the fluctuation $\phi(s)$ is assumed to be small. To
describe the energetics of the fluctuations $\phi(s)$, the bending energy term of the Hamiltonian for a worm-like chain
is augmented by a harmonic confinement potential $H_{I}$, that penalizes deviations from the circle:
$H_{I}=\frac{\lambda}{2}\int_{0}^{L}\phi^2(s)\,ds$. The harmonic potential mimics the confinement effects and the
parameter, $\lambda$, is an effective coupling constant.   This type of modeling has been successfully applied to
analyze a polymer confined in a tube \cite{Sarah, Erwin01}. The total Hamiltonian is:
\begin{align} H=\frac{\kappa}{2} \int_{0}^{L} \bigl[\frac{d\bold{u}(s)}{ds}\bigr]^2\,ds +
\frac{\lambda}{2}\int_{0}^{L}\phi^2(s)\,ds
\end{align}

Introducing two unit vectors, $\bold{e}_{\theta}(s)$ and $\bold{e}_{\rho}(s)$,
\begin{eqnarray}
\bold{e}_{\theta}(s)=(-sin\frac{s}{d},cos\frac{s}{d})\\
\bold{e}_{\rho}(s)=(cos\frac{s}{d},sin\frac{s}{d})
\end{eqnarray}
the tangent vector can be expressed as
\begin{eqnarray}
\bold{u}(s)=(1+\frac{1}{d}\phi(s))\bold{e}_{\theta}(s)+\frac{d\phi(s)}{ds}\bold{e}_{\rho}(s)
\end{eqnarray}
Then the Hamiltonian $H$ takes the following form
\begin{equation} H = \frac{\kappa L}{2d^2}+\int_{0}^{L}[(\frac{\kappa}{2d^2}+\frac{\lambda
d^2}{2})\psi^2(s)+3\kappa(\frac{d\psi(s)}{ds})^2  ~,
\end{equation} 
where $\psi(s)=\frac{\phi(s)}{d}$, the first term represents the bending energy of a circle and the
integral
represents the energy from the fluctuations. Here we make use of $\int_0^L \psi(s)\,ds = 0$ since the length of the
polymer is fixed. In addition, we have retained only the leading order gradients of $\psi$ in the harmonic
potential part in order to be consistent with the terms in the WLC model.  The
Green's function is defined as 
$G(s)=\langle\psi(s)\cdot\psi(0)\rangle$.

In the limit of $L\rightarrow\infty$ , we can decompose the fluctuations into Fourier modes,
\begin{eqnarray}
\psi(s)=\int_{-\infty}^{\infty}{e^{iqs}\psi(q)}\,dq
\end{eqnarray}
and obtain an expression for $H$ and $G(s,s^{\prime})$ in the Fourier space,respectively.
\begin{equation}
H=\frac{\kappa L}{2d^2}+\frac{L}{2\pi}\int_{0}^{\infty}[(\frac{\kappa}{2d^2}+\frac{\lambda d^2}{2}) +3\kappa
q^2]|\psi(q)|^2\,dq
\end{equation}

\begin{equation}
G(q)=\frac{2\pi}{L}\frac{1}{(\frac{\kappa}{2d^2}+\frac{\lambda d^2}{2})+3\kappa q^2}
\label{Green function}
\end{equation}
The Green's in real space is then given by $G(s) \propto e^{-s/l_{e}}$, where the length $l_{e}$ is:
\begin{equation}
\label{le} 
l_{e} = (\frac{6\kappa}{\frac{\kappa}{d^2}+\lambda d^2})^{\frac{1}{2}}
\end{equation} 
The correlation function $C(s)$ is
related to $G(s)$ by
\begin{eqnarray}
C(s)=\cos\frac{|s|}{d}(1-d^2\frac{\partial^{2}}{\partial{s}^{2}})G(s)\nonumber\\
+2d\sin\frac{|s|}{d}\frac{\partial{G(s)}}{\partial{s}}
\end{eqnarray}

Using the form of $G(s)$, we obtain:
\begin{equation}
 C(s) = e^{-s/l_{e}}(\cos\frac{|s|}{d} - 2\frac{d}{l_{e}}\sin\frac{|s|}{d} - \frac{d^{2}}{l_{e}^{2}} \cos\frac{|s|}{d}) 
\end{equation}

In the limit of small $\frac{d}{l_{e}}$ , the leading contribution to $C(s)$ is  $e^{-\frac{s}{l_{e}}}\cos\frac{s}{d}$, a
form 
that is in remarkably good agreement with the simulation results shown in Figs \ref{Fitting8}, \ref{Fitting25}, for which the condition $d <<
l_{e}$ is satisfied.  
%%Ya:  Please put in the figure labels instead of 13 and 14.  I could not easily find the labels.
The radius $d$ of the circle can be obtained by minimizing the mean-field free energy\cite{Strey}, $F$, corresponding to the Hamiltonian
$H$:
\begin{equation}
e^{-F} = \int {\cal{D}}[\psi(q)]e^{-\frac{\kappa L}{2d^2}-\frac{L}{2\pi}\int_{-\infty}^{\infty}[3\kappa
q^2+(\frac{\kappa}{2d^2}+\frac{\lambda d^2}{2})]\vert\psi(q)\vert^2\,dq}
\end{equation}
Integrating out the Gaussian functional, leads to:
\begin{equation}
 F = \frac{\kappa L}{2d^2}+\frac{L}{2\pi}\int_0^{\infty}\ln(3\kappa q^2+\frac{\kappa}{2d^2}+\frac{\lambda d^2}{2})\,dq
\end{equation}
Setting $\frac{\partial F}{\partial d} = 0$, yields
\begin{equation}
 \frac{12 \kappa^2}{\lambda d^4 -\kappa} = \sqrt{\frac{6\kappa}{\frac{k}{d^2}+\lambda d^2}}
\label{self}
\end{equation}
The effective coupling constant $\lambda$ was introduced to mimic the effects of the confining box.
Making use of the observation that $d = \frac{W}{2}$ in the simulations, we  can determine $\lambda$ in terms of $W$
and $\kappa$.   Using this result, we subsequently obtain $l_e$ in terms of $l_p$ and $W$:
\begin{equation}
 l_e = \frac{\sqrt{3}}{2}\sqrt{W^2+6l_p^2+2\sqrt{3}l_p\sqrt{3l_p^2+W^2}}
\label{leff}
\end{equation} 
With increasing values $l_{p}$, $l_{e}$ is predicted to increase linearly with $l_{p}$ for $l_{p} \gg
\frac{W}{\sqrt{6}}$. Comparing to the results of simulations (Fig.\ref{le-lp}), it its seen that the predictions of the
Gaussian theory are in semi quantitative agreement with the simulations for $l_{p} \geq 13$. This linear dependence is
different from the stiff polymer confined in the tube, for which $l_e \propto l_p^{\frac{1}{3}}$ \cite{Odijk,Sarah}.

\section{Summary} 
We have investigated the conformational and elastic properties of a single semiflexible polymer confined in 2D square box
using
numerical simulations based on a widely used lattice model of polymers, the BFM. Since effects of the competition among
configurational entropy, bending energy and excluded volume are included, this simplified model leads to non-trivial
results. By mapping the polymer configurations onto the tangent space, we visualized  changes in conformation of the
polymer as the bending rigidity was increased 
from a disordered shape to an intermediate elliptical shape and eventually to a spiral shape (Fig.
\ref{lpsmall}, \ref{lplarge}). We introduced an order parameter (Eq.
\ref{orderparameter}) to characterize the shapes and the distribution of the order parameter provided further insight
into the nature of the shape changes.
We used a meanfield approach to calculate the tangent-tangent correlation function and the
effective persistence
length.  

The combination of numerical and analytical results provides a detailed picture of the conformational and
elastic properties of a polymer under extreme confinement.   The variation of the persistence length with confinement
has been studied previously under tube confinement where the polymer is unconstrained in one dimension.   Our results
make specific predictions about the more extreme form of confinement where the polymer is constrained in all
dimensions.   The definition of the order parameter, introduced to characterize shapes in two dimensions, can be easily
generalized to higher dimensions and 
should prove useful for characterizing shapes of polymers under confinement.  
\\

\section{ACKNOWLEDGMENT} 
We acknowledge many useful discussions with Josh Kalb, Michael Hagan and Jane Kondev. This work was supported in part by
NSF-DMR-0403997.

\bibliography{Yapaper.bib}
\bibliographystyle{apsrev}

\end{document}